\documentclass[12pt,preprint]{aastex}
\usepackage{graphicx,amsmath,amssymb}
\usepackage{pdflscape}
\usepackage{url}

\newcommand{\bdv}[1]{\mbox{\boldmath$#1$}}

\def\au{{\rm AU}} 
 
\def\kms{{\rm km}\,{\rm s}^{-1}}
\def\masyr{{\rm mas}\,{\rm yr}^{-1}}

\def\muas{\mu{\rm as}}

\def\rel{{\rm rel}}

\def\eff{{\rm eff}}

\def\geo{{\rm geo}}
\def\e{{\rm E}}
\def\bpi{{\bdv\pi}}
\def\bmu{{\bdv\mu}}

\begin{document}
\title{An Earth-mass Planet in a 1-AU Orbit around an Ultracool Dwarf}

\author{Y. Shvartzvald$^{1,a}$,
J. C.\ Yee$^{2}$,
S. Calchi Novati$^{3,4}$,
A. Gould$^{5,6,7}$,
C.-U. Lee$^{5,8}$,\\
and\\
C. Beichman$^{9}$,
G. Bryden$^{1}$,
S. Carey$^{10}$,
B.~S.~Gaudi\altaffilmark{7},
C.~B.~Henderson$^{1,a}$,
W. Zhu$^{7}$\\
($Spitzer$ team)\\
and\\
M. D. Albrow$^{11}$,
S.-M. Cha$^{5,12}$,
S.-J. Chung$^{5,8}$,
C. Han$^{13}$,
K.-H. Hwang$^{5}$,
Y. K. Jung$^{2}$,
D.-J.~Kim$^{5}$,
H.-W. Kim$^{5}$,
S.-L. Kim$^{5,8}$,
Y. Lee$^{5,12}$,
B.-G. Park$^{5,8}$,
R. W. Pogge$^{7}$,
Y.-H. Ryu$^{5}$,
I.-G. Shin$^{2}$\\
(KMTNet group)}
\altaffiltext{1}{Jet Propulsion Laboratory, California Institute of Technology, 4800 Oak Grove Drive, Pasadena, CA 91109, USA}
\altaffiltext{2}{Smithsonian Astrophysical Observatory, 60 Garden St., Cambridge, MA 02138, USA}
\altaffiltext{3}{IPAC, Mail Code 100-22, Caltech, 1200 E. California Blvd., Pasadena, CA 91125, USA}
\altaffiltext{4}{Dipartimento di Fisica ``E. R. Caianiello'', Universit\`a di Salerno, Via Giovanni Paolo II, 84084 Fisciano (SA),\ Italy}
\altaffiltext{5}{Korea Astronomy and Space Science Institute, Daejon 305-348, Republic of Korea}
\altaffiltext{6}{Max-Planck-Institute for Astronomy, K\"onigstuhl 17, 69117 Heidelberg, Germany}
\altaffiltext{7}{Department of Astronomy, Ohio State University, 140 W. 18th Ave., Columbus, OH  43210, USA}
\altaffiltext{8}{Korea University of Science and Technology, 217 Gajeong-ro, Yuseong-gu, Daejeon 34113, Korea}
\altaffiltext{9}{NASA Exoplanet Science Institute, California Institute of Technology, Pasadena, CA 91125, USA}
\altaffiltext{10}{$Spitzer$, Science Center, MS 220-6, California Institute of Technology,Pasadena, CA, USA}
\altaffiltext{11}{University of Canterbury, Department of Physics and Astronomy, Private Bag 4800, Christchurch 8020, New Zealand}
\altaffiltext{12}{School of Space Research, Kyung Hee University, Yongin 446-701, Republic of Korea}
\altaffiltext{13}{Department of Physics, Chungbuk National University, Cheongju 361-763, Republic of Korea}
\altaffiltext{a}{NASA Postdoctoral Program Fellow}

\begin{abstract}
We combine $Spitzer$ and ground-based KMTNet microlensing observations
to identify and precisely measure an Earth-mass
($1.43^{+0.45}_{-0.32} M_\oplus$) planet OGLE-2016-BLG-1195Lb at
$1.16^{+0.16}_{-0.13} \au$ orbiting a $0.078^{+0.016}_{-0.012} M_\odot$
ultracool dwarf.
This is the lowest-mass  
microlensing planet to date.  At $3.91^{+0.42}_{-0.46}$ kpc, it is
the third consecutive case among the $Spitzer$ ``Galactic distribution'' planets
toward the Galactic bulge that lies in the Galactic disk as opposed
to the bulge itself, hinting at a skewed distribution of planets.
Together with previous
microlensing discoveries, the seven Earth-size planets orbiting the
ultracool dwarf TRAPPIST-1, and the detection of disks around young
brown dwarfs, OGLE-2016-BLG-1195Lb suggests that such planets might be
common around ultracool dwarfs.  It therefore sheds light on the
formation of both ultracool dwarfs and planetary systems at the limit
of low-mass protoplanetary disks.
\end{abstract}


\section{{Introduction}
\label{sec:intro}}

Formation theories suggest that planets form in
protoplanetary disks around their hosts, either through core accretion
(e.g., \citealt{Ida.2005.A}) or disk instability
(e.g., \citealt{Boss.2006.A}).  The masses and frequencies of planets
are thus tightly related to the disk masses, and indirectly to the
mass of their hosts.  Brown dwarfs (BDs) are substellar objects not
massive enough to ignite hydrogen fusion.  According to current
theory, BDs represent the lower-mass end of star formation via direct
collapse of molecular clouds (e.g., \citealt{Luhman.2012.A}).
Therefore, studying planet formation around BDs, with their low-mass
disks, probes the limiting conditions for planet formation.
\cite{Payne.2007.A} extended the study of core accretion models of
\cite{Ida.2005.A} to ultracool dwarfs (very low-mass stars and
BDs). They predict that if BDs have few-Jupiter-mass disks,
then Earth-mass planets should be frequent around them, with typical
semi-major axes $\sim 1\,\au$ and maximum planetary masses  $\sim5\,
M_\oplus$.  However, if BD disks contain only a fraction of a Jupiter
mass then they probably cannot form even an Earth-mass planet.

Early statistical studies of disks around ultracool dwarfs showed that
they are as frequent as around Sun-like stars \citep{Apai.2013.A}.
More recently, \cite{Testi.2016.A} used ALMA to study 17
young brown dwarfs and found continuum emission, indicating the
existence of dusty disks, in 11 of them.  The estimated dust masses in
these disks are  $\sim$0.5-6$M_\oplus$, suggesting total
disk masses $\sim$0.1-2$M_J$.  \cite{Daemgen.2016.A} used $Herschel$
to study 29 ultracool dwarfs and found that about half have
disks with at least one Jupiter mass.

The detection of planets around ultracool dwarfs is challenging for most planetary
detection methods because they rely on light from the faint host
and/or the even fainter planet.  Four such systems were found via
direct imaging, all with a massive planet ($>4 M_J$): 2MASS 1207-3932
\citep{Chauvin.2004.A}, 2MASS 0441-2301 \citep{Todorov.2010.A}, VHS
1256-1257 \citep{Gauza.2015.A}, and CFBDSIR 1458+1013
\citep{Liu.2011.A}. However, the companion-star mass ratios ($q>0.15$) and
large separations ($>15 \au$, except CFBDSIR 1458+1013 with $\sim$2.3
$\au$ but $q\sim0.5$) suggest they were formed similarly to binary
systems through gravitational fragmentation \citep{Lodato.2005.A}
rather than like planetary systems.
Recently, \cite{Gillon.2017.A}
detected seven terrestrial planets transiting the nearby ultracool
dwarf star TRAPPIST-1, whose host mass $0.080\pm0.007 M_\odot$
places it slightly above the hydrogen burning limit.
The planets have few-day near-resonant periods,
suggesting they formed farther out and then migrated inward.

In contrast to these discovery techniques, gravitational microlensing
does not rely on light from the system but is directly sensitive
to the masses of the planet and its host.  Its basic observable
is the Einstein timescale $t_{\rm E}$, 
which encompasses the total lens mass $M$,
the lens-source relative parallax $\pi_\rel$, and
relative proper motion $\bmu_\geo$.  Breaking the $(M,\pi_\rel,\mu_\geo)$ 
degeneracy requires
two additional parameter measurements, the angular Einstein
radius $\theta_\e$ and the amplitude of the microlens parallax vector
$\pi_\e =|\bpi_\e|$, which yield \citep{Gould.1992.B,Gould.2000.A}:
\begin{equation}
M = {\theta_\e\over\kappa \pi_\e};
\qquad
\pi_\rel = \pi_\e\theta_\e;
\qquad
\bmu_\geo = {\theta_\e\over t_\e}{\bpi_{\e,\geo}\over\pi_\e},
\label{eqn:meqn}
\end{equation}
where $\kappa\equiv {4 G\over c^2\au}\simeq 8.14{{\rm mas}\over M_\odot}$.

Four previous microlensing events revealed planets
orbiting ultracool dwarfs.  \cite{Bennett.2008.A} found a low-mass planet
in event MOA-2007-BLG-192 whose characterization was then substantially
tightened by \cite{Gould.2010.A} and \cite{Kubas.2012.A}.  The latter
used adaptive optics to measure the lens flux, from which they derived a
planet of mass  $\sim$3$\,M_\oplus$ projected
at $\sim$0.7$\au$ around a $\sim$0.08$\,M_\odot$ late M
dwarf.  However, because the planetary deviation was sparsely covered,
the uncertainties on the planet's mass and separation are large.
\cite{Gould.2014.A} discovered a $\sim$2$\,M_\oplus$ planet at
$\sim$0.8$\au$ from a 0.1-0.15$\,M_\odot$ M dwarf, which is one
member of a binary system with a slightly heavier companion
(0.12-0.18$\,M_\odot$). This planet had the lowest securely-measured
mass found by microlensing previous to the one presented here.
These two detections of planets with $\sim1\,M_\oplus$ at
$\sim 1\,\au$ around ultracool dwarfs agree well with the
predictions of \cite{Payne.2007.A}.

In addition, \cite{Han.2013.B} discovered a
$\sim$2$M_J$ planet at projected separation $\sim$0.9$\au$ orbiting
a $\sim$0.02$M_\odot$ BD.  
While the mass ratio suggests a binary-like formation mechanism,
the relatively close projected separation
hints at possible in-disk formation.  Finally,
\cite{Sumi.2016.A} discovered a planetary system in microlensing
event MOA-2013-BLG-605. They found three degenerate microlensing
models, two of which suggest a super Earth orbiting at 1--2$\au$
around a BD, while the third suggests a Neptune orbiting a
$0.2\,M_\odot$ M dwarf. The mass ratio in all solutions clearly favor
a planetary formation scenario, but the possibility of a mid-M-dwarf host
prevents the setting of strong constraints on planet formation models around BDs.

Here we analyze the planetary microlensing event
OGLE-2016-BLG-1195.  The event was observed by $Spitzer$ and several
ground surveys.  The ground light curve shows a very short anomaly
after the peak, indicating a low mass planetary companion, while the
parallax measurement from $Spitzer$ allows us to determine that it is
a $\sim$1$M_\oplus$ planet around an ultracool dwarf at
the hydrogen burning limit. The microlens modeling
yields eight degenerate solutions, which are well understood
theoretically \citep{Refsdal.1966.A,Griest.1998.A}.  However, again
for well-understood reasons, all solutions imply similar physical
properties for the system.

The $Spitzer$
microlensing campaign has enabled the systematic measurement of
the microlens parallax for well over 200 events over its first three
seasons. These will enable the first measurement of the Galactic
distribution of planets \citep{Calchi.2015.A,Zhu.2017arXiv.A}, a
demographic regime that is currently uniquely explored by microlensing
(see also \citealt{Penny.2016.A}).  The 2014 and 2015 $Spitzer$ campaigns
have already led to the publication of two other planetary systems
\citep{Udalski.2015.A,Street.2016.A}, both of which are located in the
Galactic disk, similar to OGLE-2016-BLG-1195Lb.

{\section{Observational data and reduction}
\label{sec:obs}}

OGLE-2016-BLG-1195 was alerted on UT 13:37 27 June 2016 by the
Optical Gravitational Lensing Experiment (OGLE), using the
OGLE Early Warning System \citep{Udalski.2015.B}.
The event was also observed by the Microlensing Observations
in Astrophysics (MOA) collaboration \citep{Sumi.2003.A},
which recognized and alerted the planetary anomaly 
at UT 15:45 29 June, less than two hours after it began.
The analysis and data in this paper are completely
independent from that derived from the OGLE and MOA datasets \citep{Bond.2017.A}.

\subsection{KMTNet observations}

The event was observed by Korea Microlensing Telescope Network
(KMTNet; \citealt{kmtnet}) that employs $4\,\rm deg^2$ cameras at three sites:
CTIO/Chile, SAAO/South Africa, and SSO/Australia.  At 
(RA,Dec) = (17:55:23.5, $-$30:12:26.1) (J2000.0), the event was in an
overlapping region between three high-cadence (25 min) fields, for a
combined cadence $\Gamma=7\,{\rm hr}^{-1}$, which allowed for dense
coverage of the short anomaly.  These observations were carried out as
part of the routine survey and were not influenced by the MOA alert.
Most observations were in $I$ band,
with additional sparse $V$ band observations for source
characterization.  KMTNet photometry was extracted using an adapted
version of the pySIS DIA software \citep{Albrow.2009.A} for the event
modeling, and DoPhot \citep{Schechter.1993.A} for the source color
information.

\subsection{{\it Spitzer} observations}

The event was announced by OGLE on a Monday morning (PDT), less than two
hours before the $Spitzer$ team had to submit the targets to observe
in the coming week.  Based on a preliminary model suggesting a
high-magnification event, it was selected as a ``secret'' target (see
details below).  On the following morning (PDT), the team confirmed
the model using new OGLE data, and immediately announced the event as
a ``subjective'' $Spitzer$ target on June 28 UT 15:33 (HJD'=7568.15),
about a day before the anomaly
and 3.7 days before the $Spitzer$ observations actually started.

$Spitzer$ observations are of targeted, on-going events and thus
require well defined selection criteria to avoid biases toward
planetary events.  These selection criteria are fully described in
\cite{Yee.2015.A}. Here we briefly summarize the three selection
modes.  When an event passes certain brightness and planet sensitivity
criteria (as inferred from its point-lens model) at the target
submission deadline, it is automatically selected as a $Spitzer$ target
and considered as ``objective''.  However, if an event model suggests
it {\it will} have high planet sensitivity, the team can select it 
``subjectively'', even before it meets the criteria (if it ever
does).  This allows for early $Spitzer$ observations and increases the
overall campaign sensitivity for planets while still permitting an
unbiased sample. Lastly, the team can select a promising event with a
weakly constrained point-lens model as ``secret''.  If the event
turns out to be promising (but without any indications of a planetary
perturbation), the team can select (and announce) it as
``subjective'', and it is then included in the final sample.
OGLE-2016-BLG-1195 was such event.  (If a ``secret'' event turns out to
be unpromising, it is discarded and is not included in the final
statistical sample.)

OGLE-2016-BLG-1195 was observed by $Spitzer$ during the final three
weeks of the 2016 campaign (July 2--24), with a cadence of 1
observation per day.  Each epoch is composed of six 30-second dithered
exposures using the 3.6 $\mu$m channel on IRAC,
and the data were reduced using the new algorithm for
$Spitzer$ crowded-field photometry (\citealt{Calchi.2015.B}).  The
event was faint in $Spitzer$ images, with a peak of $L\sim15.7$ and
baseline of $L\sim17.1$.  The reductions therefore required special
handling. The precise astrometric position of the source was
determined from KMTNet difference images at high magnification. 
Using this position in the analysis, the {\it Spitzer} photometry
was significantly improved relative to photometry based on the catalog
position of the apparent ``star'' at the base of the microlensing
event.

{\section{Light Curve Analysis}
\label{sec:anal}}

{\subsection{Ground-only microlensing model}
\label{sec:gblc}}

Standard binary-lens microlensing models require seven geometric parameters to
calculate the magnification, $A(t)$.  These include the three
point-lens parameters $(t_0,u_0,t_\e)$ \citep{Paczynski.1986.A}, the
angular source radius $\theta_*$ scaled to $\theta_\e$ ($\rho
= \theta_*/\theta_{\rm E}$), and three parameters for the companion:
the mass ratio $q$, the instantaneous projected separation (scaled to
$\theta_\e$) $s$, and an angle $\alpha$, between the source trajectory
and the binary axis in the lens plane.  In addition, each dataset $i$, 
has two parameters ($f_{s,i},f_{b,i}$) representing the source flux and
any additional blend flux:
\begin{equation}
f_{i}(t) = f_{s,i}A(t) + f_{b,i}.
\label{eqn:fluxpre}
\end{equation}

\begin{figure}
\plotone{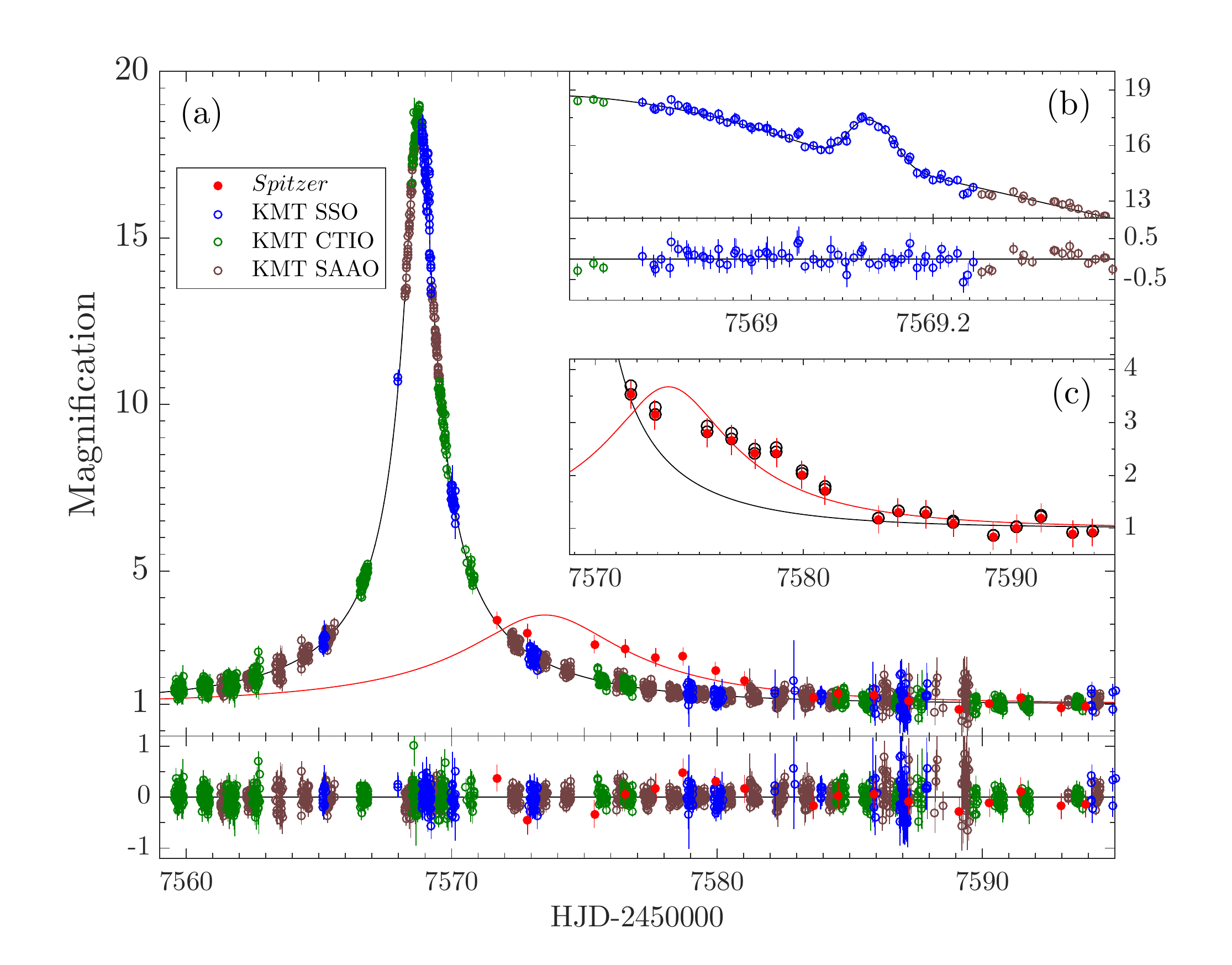}
\caption{Magnification curve of OGLE-2016-BLG-1195.
The short anomaly (see inset b) is well covered by KMT SSO.
The microlens parallax information comes from the offset between the observed $Spitzer$ points
(red circles) and the ground curve (black line, see inset c).
The open black circles are 
1$\sigma$ limits on the $Spitzer$  ``observed magnification''
 $A_{\rm ``obs''}\equiv (f_{\rm obs} - f_{\rm base})/f_s +1$,
which are independent from the parallax model.
The baseline flux $f_{\rm base}$ can be read directly off the
late-time light curve and $f_s$ is determined with 5\% uncertainty
from the $VIL$ color-color relation and $I_s$, which is
very accurately measured from the ground-based light curve.}
\label{fig:lc}
\end{figure}

The light curve (Figure~\ref{fig:lc}) has one short ($<3$ hr)
anomaly $\Delta t\simeq 8.5\,$hr after the peak of an otherwise
standard point-lens moderate-magnification ($A\simeq19$) microlensing
event. Combined with the effective timescale $t_\eff\equiv u_0 t_\e\sim 13\,$hr,
this indicates a planetary mass-ratio companion close to the Einstein ring,
aligned $\tan^{-1}13/8.5\sim 57^\circ$ relative to the source trajectory.
The single bump in the light curve implies that the source passes the single
prong of the central caustic, which faces (``points toward'') the planet
in both close and wide topologies.  We first solve for $s<1$ because this
topology is unencumbered by additional caustics on this side of the central
caustic.  The $(s>1)$ solution will immediately follow from the close-wide 
degeneracy \citep{Griest.1998.A}.

\begin{figure}
\plotone{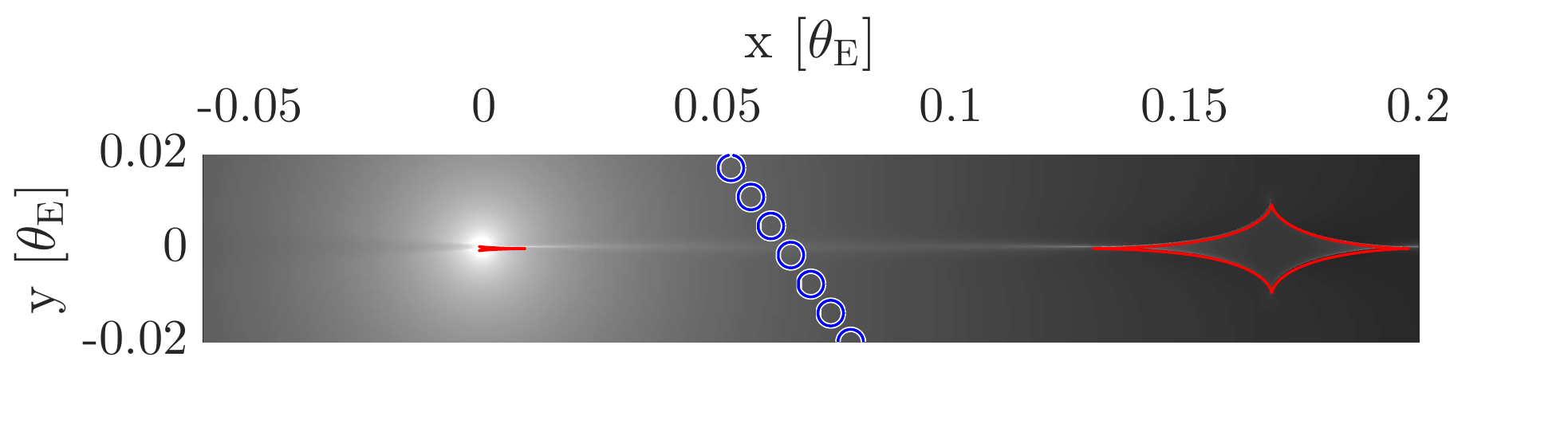}
\plotone{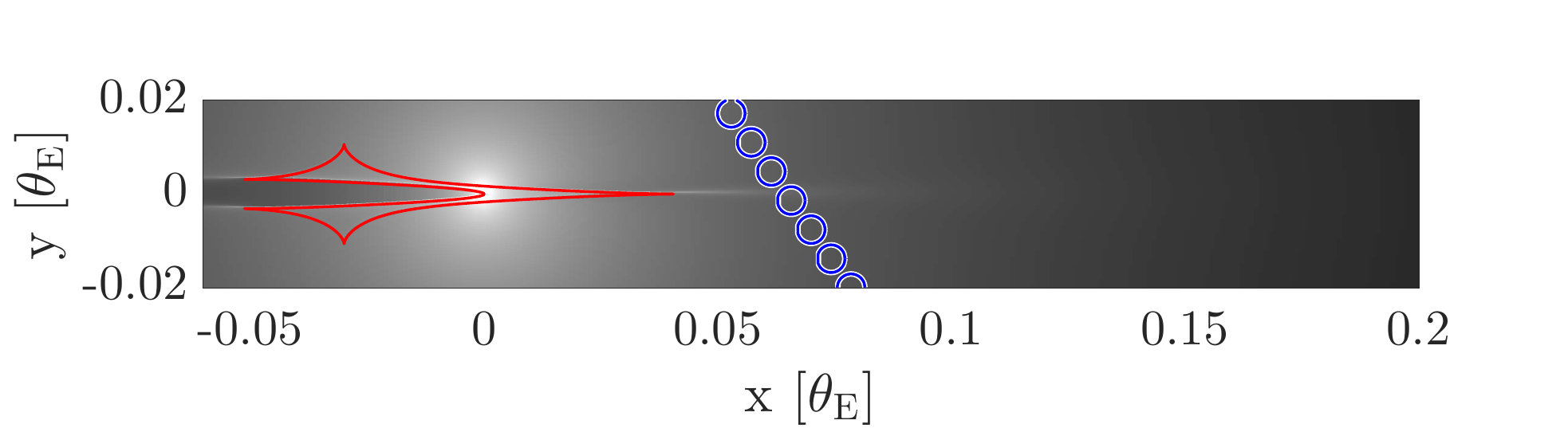}
\caption{Magnification maps of the wide ($s>1$, top) and resonant ($s<1$, bottom) topologies. The source trajectory and size are represented by the blue circles.
The red contours are the caustics. The gray scale indicates the magnification of a point source at each position, where white means higher magnification.}
\label{fig:caustics}
\end{figure}

We seed a Markov-chain Monte-Carlo (MCMC) with close 
companions ($s<1$) at an angle of $\alpha=57^{\circ}$ and with mass ratios
ranging $q=10^{-3}-10^{-6}$.
To calculate the model magnifications near and during the
anomaly, we employ contour integration \citep{Gould.1997.A} with 10 annuli
to allow for limb darkening. We adopt a linear coefficient 
$u(I) =0.526$ \citep{Claret.2000.A}, based on the source type derived in
Section~\ref{sec:cmd}.  Far from the anomaly we employ limb-darkened
multipole approximations \citep{Pejcha.2009.A,Gould.2008.A}.  
We find $(s,q)=(0.98,5.5\times 10^{-5})$, which is in the
resonant-caustic regime.  While the source does not cross the
caustic, it passes close enough and over a narrow ``shoulder'' on the
magnification surface to permit a clear measurement of the
scaled source size, $\sigma(\ln\rho)=13\%$
(Figure~\ref{fig:caustics}).

Finally, we seed the $(s > 1)$ MCMC search at $q=5.5\times10^{-5}$ and at the
boundary between the wide and resonant topologies. The result is $(s, q) =
(1.09, 5.6\times10^{-5})$ for which the source passes over a narrow ``saddle''
between two marginally separated caustics (Figure \ref{fig:caustics}).
The $(s < 1)$ and $(s > 1)$ solutions prove fully degenerate.

{\subsection{Satellite microlens parallax}
\label{sec:spitzer}}

Observations of a microlensing event by two widely separated observers
(e.g., from Earth and space) results in different observed light
curves \citep{Refsdal.1966.A,Gould.1994.A}.  Since the physical
separation between the two observers (${\bf D_\perp}$) is known,
modeling the two light curves directly yields the microlens parallax,
\begin{equation}
\bpi_\e = \frac{\au}{D_\perp}(\Delta\tau,\Delta\beta);
\qquad \Delta\tau = \frac{t_{0,\rm sat} - t_{0,\oplus} }{t_\e};
\qquad \Delta\beta = \pm u_{0,\rm sat} - \pm u_{0,\oplus},
\label{eqn:pieframe}
\end{equation}
where the subscripts indicate parameters as measured from the
satellite and Earth.  However, due to the symmetry of the problem it
usually suffers from a four-fold ``satellite'' degeneracy in
$\Delta\beta$ (Equation~(\ref{eqn:pieframe})).

When including $Spitzer$, we re-run the MCMC process with all four
possibilities for both the wide and close configurations.  In
addition, we include a constraint on the $Spitzer$ source flux,
$f_{s,Spitzer}$, derived from color-color regression (see
Section~\ref{sec:cmd}).
We find that indeed the $2\times 4=8$ possible solutions
are fully degenerate.  The magnitude 
$|\bpi_\e|$ is roughly the same for all solutions since $|u_{0,\oplus}|\ll 1$
\citep{Gould.2012.A}.
Table~\ref{tab:model} summarizes the derived model
parameters and their uncertainties for all eight solutions.

The most striking feature of these solutions is the relatively
large microlens parallax $\pi_\e\sim 0.45$, which (combined with
the small value of $\theta_\e$) will yield small masses for both
the host and the planet (see Section \ref{sec:earth}). However,
Figure~\ref{fig:lc} demonstrates, independent of any model, that there is a
significant parallax effect. First, inset (c) shows that the {\it Spitzer}
magnification is larger than the ground-based magnification at the same
time. Because the peak ground-based magnification is quite high
$A_{\rm max,\oplus}\simeq u_{0,\oplus}^{-1}\simeq 19$,  {\it Spitzer}'s
substantially higher post-peak magnification directly implies that
it peaks substantially later.  The fact that the first {\it Spitzer}
points nearly overlap the ground-based magnification then implies substantially
larger $u_{0,Spitzer}$ as well.
One may therefore qualitatively infer 
that both components of the parallax should be large.

To make this argument more concrete, we consider a range of models
with different $t_{0,Spitzer}=7569, 7570, \ldots 7575$, and with
the {\it Spitzer} flux constraint set at $-2\,\sigma$, $0\,\sigma$,
and $+2\,\sigma$ from value that we derive from the $VIL$ color-color
relation.  The lowest of these value $t_{0,Spitzer}=7569$ would yield
the lowest $|\Delta t_0|$ and therefore the lowest value of
$|\pi_{\e,E}|$ (and so potentially the lowest value of $\pi_\e$,
depending on the value of $\Delta \beta$ for that solution).
This (and other low $t_{0,Spitzer}$) solutions also seem naively in
accord with the fact that the data appear to fall almost monotonically
over the first nine points, whereas the best model peaks between the second
and third data points.  The last value in this series is a few days
later than in the best fit model.

\begin{figure}
\plotone{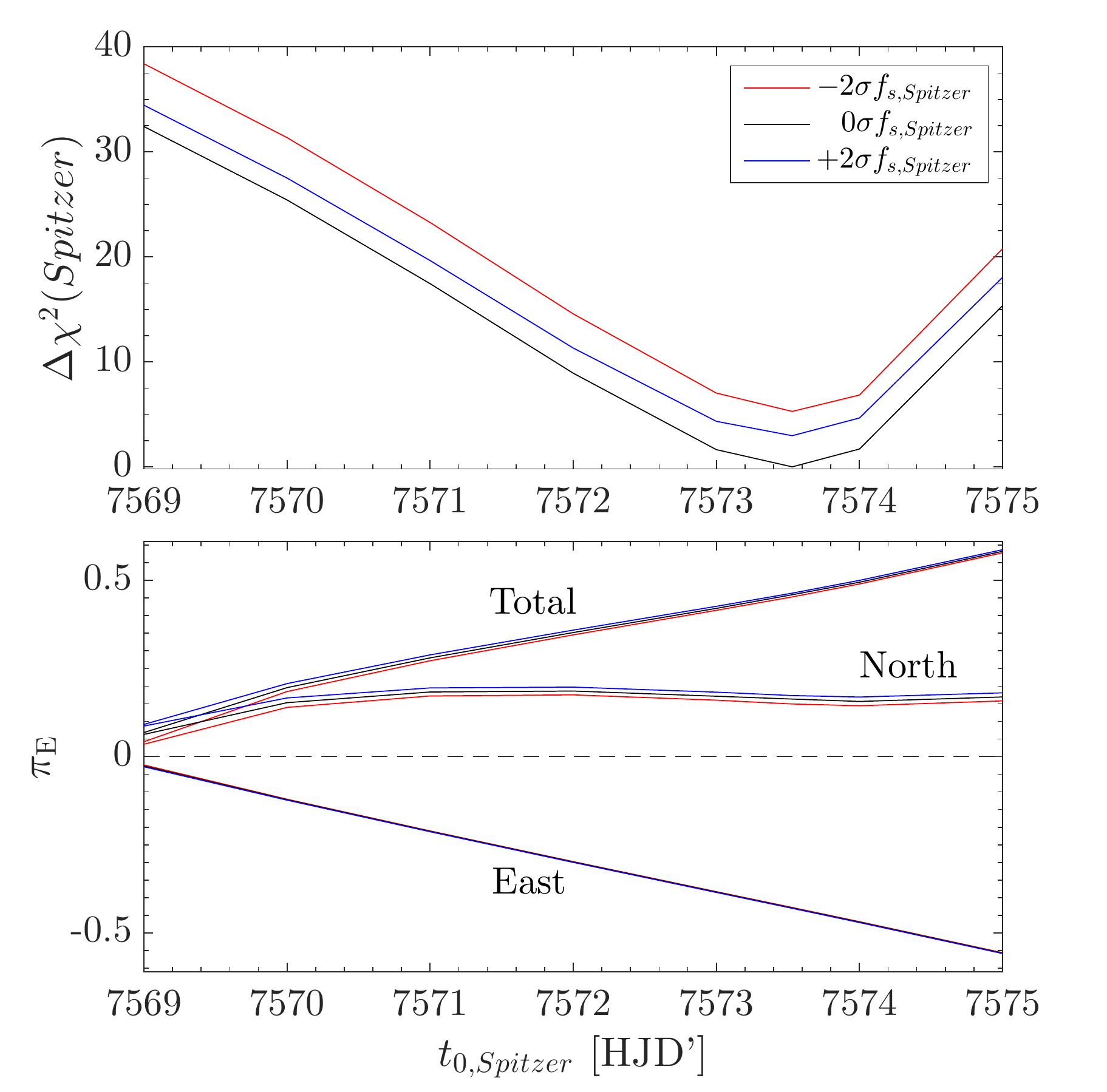}
\caption{Test of robustness of microlens parallax $\bpi_\e$ measurement.
Lower panel shows values of $\pi_{\e,\rm E}$ (bottom), $\pi_{\e,\rm N}$ (middle)
and total $\pi_\e$ (top) for the close($++$) solution when $t_{0,Spitzer}$ is forced to different values
(abscissa). Red, black and blue tracks show respectively the $(-2,0,+2)\sigma$
deviations from the best estimate of $f_{s,Spitzer}$ derived from the $VIL$
color-color constraint.  These are barely distinguishable.
Other ($++$) and ($--$) solutions are extremely similar and
the ($+-$) and ($-+$) solutions are broadly similar but with
much larger parallax at low $t_{0,Spitzer}$ (as expected from their geometry).
The upper panel
displays $\Delta\chi^2$ relative to the best fit.  The figure shows that
$\pi_\e$ can be forced down substantially only by forcing $t_{0,Spitzer}<7571$,
where $\Delta\chi^2$ is already very high. 
See text for physical explanation
of this robustness.
}
\label{fig:forcedt0}
\end{figure}

The results of this test are shown in Figure~\ref{fig:forcedt0}.  The lower
panel demonstrates that indeed if $t_{0,Spitzer}$ is artificially pushed
lower, the value of $\pi_\e$ is reduced.  However, for
$t_{0,Spitzer}\geq 7571$, this is almost entirely due to shrinking 
$|\pi_{\e,E}|$, while $\pi_{\e,N}$ remains essentially constant.
Hence, at this point $\pi_\e$ is still 60\% of the best-fit value
(meaning that the main physical conclusions would be qualitatively
similar), while $\Delta\chi^2=17.5$ is already quite high.  Note
also that the $\bpi_\e$ trajectories for different $f_{s,Spitzer}$
(shown by different colors) are essentially identical, with the
major difference being the higher $\Delta\chi^2= 4$ penalty 
for the $\pm 2\sigma$ tracks.  

The basic reason that $\pi_\e$ changes very little over this range
is that, a week after the ground-based peak 
$(\tau,\beta)_\oplus = (0.7,0.05)$, while the ``observed'' {\it Spitzer}
magnification is
$A_{Spitzer}=3$, i.e., $u_{Spitzer}=0.35$.  This implies $|\Delta{\bf u}|>0.35$
(hence $\pi_\e> (\au/D_\perp)|\Delta{\bf u}|=0.30$) unless the model
is forced to systematically fall below the {\it Spitzer} data.

{\section{Physical properties}
\label{sec:physical}}

The mass and distance of the lensing system can be derived from
$\pi_\e$ and $\theta_\e$ (Equation~\ref{eqn:meqn}).  
These allow us to use the mass ratio $q$ and
the scaled projected separation $s$ to derive 
the masses and physical projected separation of the two bodies.
While $\pi_\e$ is directly measured in the microlensing
model, $\theta_\e$ is derived from $\rho$
(found from the light curve model) and $\theta_*$
(found using the color-magnitude diagram (CMD)): $\theta_\e=\theta_*/\rho$.

{\subsection{CMD}
\label{sec:cmd}}

\begin{figure}
\plotone{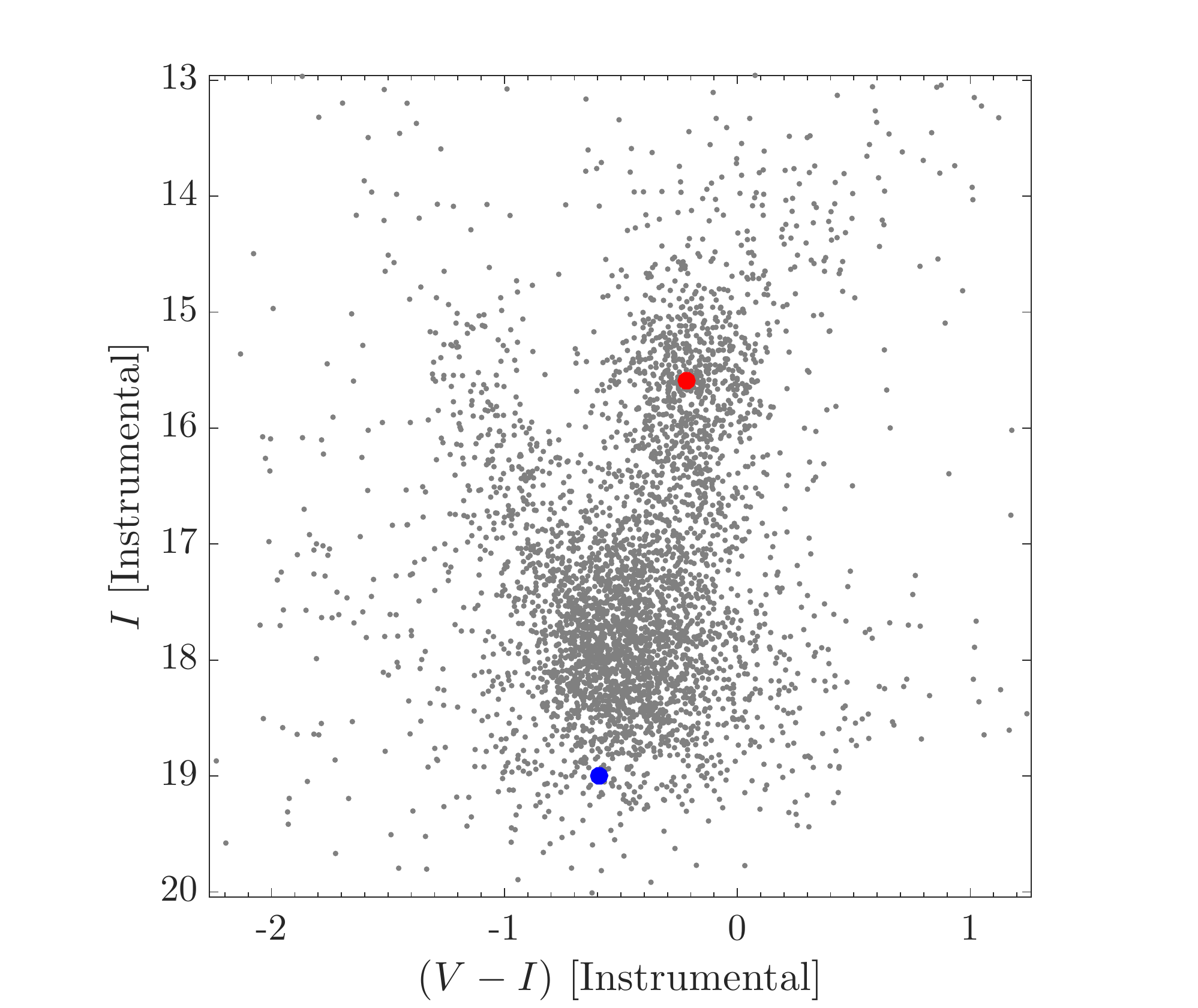}
\caption{KMTNet instrumental CMD of stars 
from a $205''\times205''$ field centered on
the event's position.
The offset between the red clump centroid (red) and the source star (blue) allows us to derive the source angular radius $\theta_*$.
}
\label{fig:cmd}
\end{figure}

The source properties are derived from its position on a CMD, constructed
using stars from a $205''\times205''$ field centered on
the event's position
(Figure~\ref{fig:cmd}), using KMTNet instrumental $V$-band and $I$-band
magnitudes.  We measure the centroid of the ``red giant clump'' (RGC)
$(V-I,I)_{\rm cl, kmt}=(-0.22 ,15.59)$ and compare it to the
intrinsic centroid of $(V-I,I)_{\rm cl,0}=(1.06,14.44)$ 
\citep{Bensby.2013.A,Nataf.2013.A} for the event's Galactic
coordinates $(l,b)=(0.0,-2.5)$.  The KMTNet $I$-band source
magnitude as inferred from the microlensing model is 
$I_{s, {\rm kmt}}=18.99\pm 0.02$, and assuming it is behind the same dust
column as the red clump, its intrinsic magnitude is $I_{s,0}=17.84\pm
0.04$.

A standard way to determine the instrumental $(V-I)_s$ color is from
regression of $V$ versus $I$ flux as the source magnification changes
\citep{Gould.2010.A}. Applying this to KMTNet data yields
$(V-I)_{s, {\rm  kmt}}=-0.59\pm0.02$, from which we derive 
$(V-I)_{s,0}=0.68\pm 0.03$ by correcting to the clump offset found above.
Using standard color-color relations \citep{Bessell.1988.A} and
the relation between angular source size and surface brightness
\citep{Kervella.2004.A}, we find 
$\theta_*=0.82\pm 0.07\,\muas$.

We use red giant branch stars 
($14.6<I_{\rm KMT}<16.6\,;\,-0.6<(V-I)_{\rm KMT}<0.2$), 
which are a good representation of the bulge star population, 
to derive an instrumental $VIL$ color-color relation.
Since the source is significantly bluer than this range we apply a new color-color method.
For each giant we use the dereddened $(V-I)_0$ color to find 
the corresponding $(I-L)_0$ based on \cite{Bessell.1988.A} giants,
and derive the offset between the instrumental ($I_{\rm KMT}-L_{Spitzer}$)
and the $(I-L)_0$ for the sample.
We then use the source $(V-I)_{s,0}$ color and find its corresponding $(I-L)_{s,0}$
based on \cite{Bessell.1988.A} dwarfs\footnote{Note that while the 
\citet{Bessell.1988.A} $VIL$ relations are derived for ground-based $L$ band,
which differs somewhat from the {\it Spitzer} $3.6\,\mu$m band, it can
still be used essentially unmodified for our purposes.  This is because
these authors report $(K-L)\propto 0.04(V-K)$, which is already quite small,
and the further correction from $L$ to $3.6\,\mu$m is an order of
magnitude smaller.}. Finally, we use the color offset
to find the source instrumental color, $(I_{\rm KMT}-L_{Spitzer})_s=0.78\pm 0.03$,
which is used to constrain the {\it Spitzer} source-star flux in the modeling.
For comparison, we also derive the source instrumental color using linear
$VIL$ color-color relation using the giants and extrapolate it to the source
(see \citealt{Calchi.2015.B} for didactic explanation).
The color we find using this method is $(I_{\rm KMT}-L_{Spitzer})_s=0.77\pm 0.05$,
which is in agreement with the new method.
While the previous method works extremely well for colors in the range
of the giants, it is important to use the new method when extrapolating
from the calibration range.

{\subsection{An Earth-mass planet around an ultracool dwarf}
\label{sec:earth}}

As mentioned above, the four-fold degenerate parallax solutions ($\bpi_\e$) have approximately
the same magnitude, and so give similar masses, distance from Earth
and relative proper motion of the planetary system.  In addition, the
wide and close solutions are very close to $s=1$, and so the projected
separation is also similar.  In summary, the derived physical
properties of the eight degenerate model solutions are all in agreement
within 1$\sigma$.  Therefore, we combine the results from the
different models and give the median and 68\% uncertainty range of the
physical properties of the system in Table~\ref{tab:physical}.

The host star mass is $0.078^{+0.016}_{-0.012} M_\odot$ (or
$81^{+17}_{-13} M_J$), i.e., with mean just at the hydrogen-burning mass
limit at Solar metallicity.  The companion mass is
$1.43^{+0.45}_{-0.32} M_\oplus$, which is the lowest mass planet
discovered by microlensing, and it orbits the ultracool dwarf at a projected
separation of $1.16^{+0.16}_{-0.13} \au$.  The system lies in the
Galactic disk, at $3.91^{+0.42}_{-0.46}$ kpc toward the bulge.

The relative proper motion between the source and the lens is
$\mu_{\rm hel}=8.7^{+1.6}_{-1.2}$ mas/yr, which is consistent with
expectations for disk lenses.  However, the direction of the proper
motion is clustered at 
$\mu_{\rm hel}{\rm(N,E)}\approx(\pm4.0,-7.5)\,\masyr$, whereas
typical disk lenses at $D_L\sim 4\,$kpc would have 
$\mu_{\rm hel, E}\sim +3\,\masyr$.  This discrepancy could be
resolved by the lens moving $\sim 180\,\kms$ relative to its
local standard of rest (i.e., halo lens) or the source moving
$\sim 360\,\kms$ relative to the bulge (or some combination).
Both possibilities would seem to have low prior probability.
However, recall that in Section~\ref{sec:spitzer} we showed
that the negative value of $\pi_{\e,\rm E}$ (hence $\mu_{\rm E}$) is a direct
consequence of the relatively high magnification as seen by {\it Spitzer}
one week after $t_{0,\oplus}$, which is attested to by a whole series of
{\it Spitzer} measurements.  Hence, it is not easily avoided.
The issue can be resolved by two epochs of
high resolution imaging, which would measure the source proper motion
relative to the bulge, $\bmu_{S}$, and so yield directly the lens
proper motion $\bmu_{L} = \bmu_{S} + \bmu_{\rm hel,rel}$.
  
{\section{Discussion}
\label{sec:discussion}}

We have detected and characterized an Earth-mass planet orbiting an
ultracool dwarf (at the 
hydrogen-burning limit, assuming Solar metallicity)
at $\sim$1$\,\au$. This system adds to two previous microlensing
discoveries of planets of a few Earth masses orbiting ultracool
dwarfs. These suggest that the protoplanetary disks of ultracool dwarfs
have sufficient mass to form terrestrial planets, as also hinted at by
direct imaging of such disks. The location of these planets, at about
$1\,\au$, support planet formation predictions. However, since the
sensitivity of current microlensing surveys for systems with such
small mass ratios is very narrow, around projected separations of
$\sim$1$\au$, they cannot set strong constraints on the presence of
planets elsewhere around ultracool dwarfs, such as the much closer separations
seen in the TRAPPIST-1 system.  

The dense coverage of the short duration
anomaly induced by the low mass-ratio planet OGLE-2016-BLG-1195Lb was
enabled due to the high-cadence coverage of KMTNet. This shows the
importance of the new high-cadence global network surveys, which do
not require the traditional alert and follow up mode, previously used
in microlensing.  Future space-based microlensing surveys, such as
$WFIRST$ \citep{Spergel.2015arXiv.A}, will have this
required high-cadence and with their superior photometry will extend
the detection sensitivity to wider separations and lower planetary
masses.

This is the third planet discovered as part of the $Spitzer$
microlensing campaign for measuring the Galactic distribution of
planets, and another few planetary candidates from the 2015 and 2016
seasons are currently being investigated. So far, all of the
published planets are located in the Galactic disk.  \cite{Zhu.2017arXiv.A}
studied the planet sensitivities of 2015 $Spitzer$
events with high-cadence KMTNet coverage
and found that if the frequency of planets is equal in the
bulge and in the disk, about 1/3 of the planet detections should be
from systems in the bulge.  With two final $Spitzer$ microlensing
campaigns in 2017 and 2018, and the expected additional planetary
detections, we will be able to check for deviations from
this expectation, and see if the bulge is deficient of
planets, as the current detections hints and as suggested by
\cite{Penny.2016.A}.

\acknowledgments
We thank D. Kirkpatrick for fruitful discussions about BDs.
This work is based in part
on observations made with the {\it Spitzer} Space Telescope,
which is operated by the Jet Propulsion Laboratory, California Institute of Technology under a contract with
NASA.
This research has made use of the KMTNet system operated by the Korea
Astronomy and Space Science Institute (KASI) and the data were obtained at
three host sites of CTIO in Chile, SAAO in South Africa, and SSO in
Australia.
Work by YS and CBH was supported by an
appointment to the NASA Postdoctoral Program at the Jet
Propulsion Laboratory, administered by Universities Space Research Association
through a contract with NASA.
Work by AG, JCY and SCN were supported by JPL grant 1500811.
Work by AG, WZ, YKJ and IGS were supported by NSF grant AST-1516842.
Work by C.H. was supported by Creative Research Initiative Program
(2009-0081561) of National Research Foundation of Korea. 
Work by SCN was supported by NExScI.
Copyright 2017. All rights reserved.

\begin{landscape}
\begin{table}
\tiny
\centering
\caption{Best-fit microlensing model parameters
and their 68\% uncertainty range derived from the MCMC chain density
for the eight degenerate solutions \label{tab:model}}
\begin{tabular}{l|cccc|cccc}
\tableline\tableline
Parameter	& \multicolumn{4}{c}{Close}	& \multicolumn{4}{c}{Wide}	\\
& $--$	& $-+$	& $+-$	& $++$	& $--$	& $-+$	& $+-$	& $++$	\\
\tableline\\
$t_0$ [HJD']& $7568.7692\pm0.0013$& $7568.7695\pm0.0013$& $7568.7693\pm0.0013$& $7568.7693\pm0.0013$& $7568.7694\pm0.0013$& $7568.7695\pm0.0013$& $7568.7694\pm0.0013$& $7568.7695\pm0.0012$\\[5pt]
$u_0$& $-0.05321\pm0.00073$& $-0.05284\pm0.00079$& $0.05317\pm0.00075$& $0.05320_\pm0.00073$& $-0.05330\pm0.00073$& $-0.05321\pm0.00075$& $0.05321\pm0.00074$& $0.05324\pm0.00074$\\[5pt]
$t_{\rm E}$ [d]& $9.96\pm0.11$& $10.01\pm0.11$& $9.96\pm0.11$& $9.96\pm0.11$& $9.94\pm0.11$& $9.95\pm0.11$& $9.95\pm0.11$& $9.95\pm0.11$\\[5pt]
$\rho$[$10^{-3}$]& $2.90_{-0.40}^{+0.34}$& $2.87_{-0.41}^{+0.35}$& $2.89_{-0.38}^{+0.34}$& $2.90_{-0.40}^{+0.34}$& $2.86_{-0.43}^{+0.34}$& $2.85_{-0.42}^{+0.35}$& $2.84_{-0.42}^{+0.35}$& $2.85_{-0.41}^{+0.35}$\\[5pt]
$\pi_{\rm E,N}$& $-0.2154\pm0.0065$& $0.2335\pm0.0080$& $-0.3017\pm0.0071$& $0.1487\pm0.0074$& $-0.2158\pm0.0066$& $0.2350\pm0.0080$& $-0.3016\pm0.0074$& $0.1491\pm0.0075$\\[5pt]
$\pi_{\rm E,E}$& $-0.380\pm0.032$& $-0.411\pm0.032$& $-0.376\pm0.032$& $-0.404\pm0.031$& $-0.382\pm0.032$& $-0.413\pm0.031$& $-0.377\pm0.032$& $-0.404\pm0.032$\\[5pt]
$\alpha$ [rad]& $-0.9684\pm0.0022$& $-0.9681\pm0.0022$& $0.9684\pm0.0022$& $0.9684\pm0.0022$& $-0.9690\pm0.0022$& $-0.9688\pm0.0022$& $0.9688\pm0.0022$& $0.9689\pm0.0022$\\[5pt]
$s$& $0.9842_{-0.0075}^{+0.0069}$& $0.9834_{-0.0072}^{+0.0070}$& $0.9839_{-0.0072}^{+0.0068}$& $0.9840_{-0.0075}^{+0.0068}$& $1.0856_{-0.0075}^{+0.0082}$& $1.0861_{-0.0076}^{+0.0081}$& $1.0862_{-0.0075}^{+0.0080}$& $1.0853_{-0.0074}^{+0.0084}$\\[5pt]
$q$ [$10^{-5}$]& $5.43_{-0.70}^{+0.82}$& $5.49_{-0.73}^{+0.78}$& $5.46_{-0.71}^{+0.81}$& $5.47_{-0.72}^{+0.79}$& $5.60_{-0.70}^{+0.86}$& $5.68_{-0.73}^{+0.83}$& $5.68_{-0.72}^{+0.80}$& $5.58_{-0.71}^{+0.84}$\\[5pt]
$\chi^2$& 10214& 10214& 10213& 10213& 10214& 10214& 10213& 10214\\[5pt]
\tableline\tableline
\end{tabular}
\newline
\raggedright{HJD'=HJD-2450000}
\end{table}
\end{landscape}

\begin{table}
\centering
\caption{Physical properties of the planetary system \label{tab:physical}}
\begin{tabular}{l|c}
\tableline\tableline
$M_1$ [$M_\odot$]&  $0.078^{+0.016}_{-0.012}$\\[5pt]
$M_2$ [$M_\oplus$]&  $1.43^{+0.45}_{-0.32}$\\[5pt]
$r_\perp$\,\, [$\au$]&  $1.16^{+0.16}_{-0.13}$\\[5pt]
$D_L$ [kpc]&  $3.91^{+0.42}_{-0.46}$\\[5pt]
$\theta_E$ [mas]&  $0.286^{+0.053}_{-0.038}$\\[5pt]
$\mu_{\rm hel}$ [mas/yr]&  $8.7^{+1.6}_{-1.2}$\\[5pt]
\tableline\tableline
\end{tabular}
\end{table}

\end{document}